\documentclass[10pt, twocolumn, journal]{IEEEtran}
\usepackage{amsmath}
\usepackage{algorithm}
\usepackage{algorithmic}
\usepackage{cite}
\usepackage{makecell}
\usepackage{url}
\usepackage{breakurl}
\usepackage{slashbox}
\usepackage[breaklinks]{hyperref}
\usepackage{epsfig,psfig}
\usepackage{graphicx}
\usepackage{dblfloatfix}
\usepackage{multirow}
\usepackage{threeparttable}
\usepackage{adjustbox}
\usepackage{booktabs}
\usepackage{subcaption}
\usepackage{verbatim}

\begin{document}

\title{Toward E2E Intelligence in 6G Networks: An AI Agent-Based RAN-CN Converged Intelligence Framework}

\author{Youbin Han, Haneul Ko,~\IEEEmembership{Senior Member,~IEEE,} Namseok Ko, Tarik Taleb,~\IEEEmembership{Senior Member,~IEEE,} and Yan Chen

\thanks{Y. Han and H. Ko are with Kyung Hee University, Korea. (e-mail: \{gksyb4235,heko\}@khu.ac.kr), corresponding authors: H. Ko).}

\thanks{N. Ko is with the Mobile Core Network Research Section, Electronics and Telecommunications Research Institute, Daejeon, Korea. (e-mail: nsko@etri.re.kr).}

\thanks{T. Taleb and Y. Chen are with Ruhr University Bochum, Germany. (e-mail: \{Tarik.Taleb,Yan.Chen-j3l\}@ruhr-uni-bochum.de).}

\thanks{This work was supported in part by National Research Foundation (NRF) of Korea Grant funded by the Korean Government (MSIP) (No. RS-2024-00340698) and by the ICT R\&D program of MSICT/IITP. [RS-2024-00405354, Development of Evolved SBA Framework and Core Technologies of Control/User Plane NFs].}

}

\maketitle

%{Shell \MakeLowercase{\textit{et al.}}: Bare Demo of IEEEtran.cls for Computer Society Journals}

\begin{abstract}
Recent advances in intelligent network control have primarily relied on task-specific Artificial Intelligence (AI) models deployed separately within the Radio Access Network (RAN) and Core Network (CN). While effective for isolated models, these suffer from limited generalization, fragmented decision-making across network domains, and significant maintenance overhead due to frequent retraining. To address these limitations, we propose a novel AI agent-based RAN-CN converged intelligence framework that leverages a Large Language Model (LLM) integrated with the Reasoning \& Acting (ReAct) paradigm. The proposed framework enables the AI agent to iteratively reason over real-time, cross-domain state information stored in a centralized monitoring database and to synthesize adaptive control policies through a closed-loop thought–action–observation process. Unlike conventional Machine Learning (ML) based approaches, it does not rely on model retraining. Instead, the AI agent dynamically queries and interprets structured network data to generate context-aware control decisions, allowing for fast and flexible adaptation to changing network conditions. Experimental results demonstrate the enhanced generalization capability and superior adaptability of the proposed framework to previously unseen network scenarios, highlighting its potential as a unified control intelligence for next-generation networks.
\end{abstract}


\section{Introduction}

As mobile networks evolve toward 6G, intelligence is rapidly becoming a foundational capability. Standardization bodies such as 3rd Generation Partnership Project (3GPP) and the Open Radio Access Network (O-RAN) Alliance are accelerating Artificial Intelligence (AI) integration across both the Radio Access Network (RAN) and Core Network (CN)~\cite{TR22.850, TR23.700-84, TR28.866, TS23.288, TS28.105, O-RAN}. For instance, the Radio Intelligence Controller (RIC) deploys AI-based xApps to optimize radio behaviors including resource allocation, mobility decisions, and interference control, while the Network Data Analytics Function (NWDAF) in the CN provides AI-driven analytics for traffic prediction, anomaly detection, and capacity planning.

Despite this progress, today's AI-based network intelligence still faces fundamental structural limitations. First, most deployed models are inherently task-specific and operate independently within their respective domains~\cite{TR22.850, TR23.700-84}. Each model optimizes a narrowly defined objective and operates in isolation, without explicit mechanisms for coordinating decisions across the RAN and CN. Consequently, this domain isolation leads to fragmented reasoning and inconsistent actions—for example, the RIC may trigger unnecessary handovers even after the CN has already scaled User Plane Function (UPF) capacity to alleviate congestion.

Second, conventional Machine Learning (ML) models struggle to generalize to unseen environments because they rely on specific training datasets and fixed input-output mappings. When traffic patterns shift, user behavior changes, or deployment conditions differ from the training distribution, model performance deteriorates sharply. This degradation necessitates frequent retraining, increasing operational overhead and limiting real-time adaptability~\cite{TS28.105}.

Third, existing architectures lack a unified reasoning framework capable of jointly interpreting RAN and CN state information~\cite{TR28.866, TS23.288, O-RAN}. As a result, networks cannot diagnose cross-domain causal relationships—for instance, whether radio congestion is the root cause of downstream CN overload, or whether CN-level resource scarcity drives repeated signaling failures at the RAN. Without integrated reasoning across layers, true End-to-End (E2E) optimization becomes unattainable.

To overcome these limitations, next-generation network intelligence must reason across domains, maintain adaptability without retraining, and generate decisions that reflect real-time cross-domain context. Recent advances in Large Language Models (LLMs) and AI agent architectures offer a promising pathway~\cite{Zhu25}. Unlike task-specific ML models, LLMs can process heterogeneous data—including telemetry metrics, logs, KPIs, and contextual information—enabling unified reasoning across the RAN and CN and directly addressing the fragmentation caused by siloed models.

LLMs also possess strong generalization capability, as their reasoning is not tied to fixed mappings learned from a single dataset. By leveraging broad prior knowledge and semantic understanding, they infer solutions even under previously unseen scenarios without retraining. Moreover, their ability to analyze multiple dimensions of network state allows them to identify cause-effect relationships between RAN and CN behaviors, a capability that conventional domain-specific ML models lack~\cite{Shahid25}.

When augmented with the Reasoning \& Acting (ReAct) paradigm, LLMs evolve from passive inference engines into active, iterative decision-makers~\cite{ReAct}. ReAct enables the agent to gather missing information, query a real-time monitoring database (DB), validate hypotheses, and refine decisions before generating control actions. This creates a closed-loop reasoning cycle, allowing the agent to continuously align its policies with evolving RAN-CN conditions and respond in a context-aware, adaptive manner.

This article introduces an AI agent-based RAN-CN converged intelligence framework that leverages LLM-driven reasoning with the ReAct paradigm to unify RAN and CN decision processes. The framework connects the agent to a real-time monitoring DB that aggregates cross-domain state information—such as radio resource utilization, user density distributions, UPF delay metrics, and slice-level Quality of Service (QoS) violations. Through iterative reasoning, the agent issues structured queries, refines its internal understanding, and generates high-level policies aligned with current conditions. Evaluation results demonstrate that the proposed framework enables unified reasoning across multiple network tasks and improves E2E slice optimization for adaptive and coherent 6G network intelligence.

The key contributions of this article are as follows:
\begin{itemize}
    \item We analyze architectural constraints of current RAN and CN intelligence, showing how task-specific ML models, poor generalization, and isolated control domains limit E2E optimization.
    \item We propose a unified AI agent architecture based on LLMs and the ReAct paradigm, enabling iterative reasoning, dynamic querying, and cross-domain policy generation aligned with real-time network state.
    \item We demonstrate the agent's adaptability and generalization capability through multi-task predictions and end-to-end network slicing optimization under diverse and previously unseen environments.
    \item We explore future directions toward AI-native 6G systems, including optimized agent deployment, safe ReAct control, multi-agent intelligence, intelligent data pipelines, and standards-compliant validation.
\end{itemize}

The remainder of this article is organized as follows. Section II discusses the key structural limitations of existing AI/ML-based network intelligence. Section III reviews related work and background on AI agent architecture. Section IV presents the AI agent-based RAN-CN converged intelligence framework. Section V evaluates the performance of the proposed framework across various scenarios. Section VI outlines future research directions, and Section VII concludes the paper.

\section{Key Limitations}

Despite significant advances in integrating AI into mobile networks, current network intelligence remains constrained by task-specific models, limited generalization, and RAN-CN domain isolation. These challenges prevent holistic, adaptive, and cross-domain intelligence required for 6G. This section explains these three structural limitations and their impact on E2E intelligence.

\textbf{Limitations of task-specific ML models:} Current AI models in mobile networks are highly specialized, addressing specific tasks such as radio resource allocation, mobility prediction, anomaly detection, or UPF load estimation. While effective within narrow scopes, they lack the flexibility to capture broader system dynamics or evolving context across different network domains. This specialization creates operational overhead, as each model requires a separate data pipeline, training, and validation. As deployments expand, operators must manage multiple model lifecycles in parallel, increasing the Operations, Administration, and Maintenance (OAM) burden and complicating coordinated intelligence~\cite{Shi23}. Even small updates or retraining events can propagate inconsistencies across subsystems, revealing scalability challenges in model-centric architectures.

\textbf{Poor generalization to unseen scenarios:} Traditional AI models are typically trained on static or geographically limited datasets. However, live mobile networks experience continuous variability driven by mobility, fluctuating user density, evolving service mixes, device heterogeneity, and unforeseen faults. When real-world conditions deviate from the training distribution, performance often degrades sharply~\cite{Akrout23}. This issue is particularly impactful for time-sensitive functions such as traffic prediction, congestion handling, and mobility management, where degradation can cascade across the network. Restoring performance requires expensive, time-consuming retraining or fine-tuning, which limits real-time adaptability and stable operation under evolving network conditions.

\textbf{Lack of a unified reasoning framework for cross-domain optimization:} A fundamental obstacle stems from the domain-isolated design of current AI control architectures. Without an integrated reasoning layer, systems fail to capture cross-domain interactions, often leading to misaligned decisions. For example, when NWDAF initiates UPF scaling to mitigate packet-processing delays, the RIC—unaware of this update—may still execute unnecessary handovers or radio resource reallocation based on outdated state information. Conversely, RAN congestion can be misread by the CN as registration failure, resulting in repeated session attempts and excessive signaling overhead. Such miscoordination undermines QoS assurance and slice performance. Advanced functions like Service Level Agreement (SLA)-aware resource allocation or cross-domain mobility control inherently require unified interpretation of radio- and core-level conditions, which is fundamentally absent in current architectures.

\section{Related Works \& Background}
\label{Sec:Related_works}

\subsection{Related Works for E2E Intelligence}

To overcome the structural fragmentation of current network intelligence, the evolution toward E2E intelligence has become a primary focus for 6G. This paradigm entails the native embedding of intelligence across all network segments to facilitate seamless cross-domain coordination and holistic system optimization. This section reviews the standardization and technical research that bridge the gap between task-specific models operating in isolation and cross-domain AI agents reasoning in a unified manner.

\textbf{Standard activities:} 3GPP has launched several initiatives to reduce fragmentation in AI/ML adoption across network domains. One representative effort is the study on AI/ML consistency alignment~\cite{TR22.850}, which seeks to identify and mitigate inconsistencies in AI/ML applications across the RAN, CN, and Service and System Aspects (SA) domains. Through these efforts, 3GPP seeks to align domain-specific intelligence and enable consistent E2E reasoning across network layers.

In parallel, 3GPP is advancing the cross-domain AI/ML collaboration~\cite{TR23.700-84} to support more integrated network control. One key focus is the enhancement of location-based services by integrating RAN-side radio measurements with CN-side analytics from NWDAF. This highlights the need for new data collection structures and extended measurement capabilities, particularly within the RAN Working Groups (WGs).

Beyond functional collaboration, 3GPP is also exploring the operational and management aspects of AI/ML deployment, including AI/ML lifecycle management~\cite{TS28.105}, AI-powered analytics for network management and orchestration~\cite{TR28.866}, and enhanced NWDAF functions for real-time decision-making~\cite{TS23.288}. These efforts are coordinated with key WGs, such as RAN WG3, SA WG2, and the ETSI Industry Specification Group on Zero-Touch Network and Service Management (ZSM), to improve interoperability across standards.

Another key challenge for E2E intelligence is transforming the RAN into a Service-Based Architecture (SBA). RAN-CN convergence remains technically difficult due to the strong coupling and hierarchical structure of RAN protocols. The core issue lies in decomposing and reconstructing the layered RAN protocol stack within the SBA paradigm. In this context, the O-RAN Alliance proposes disaggregating RAN functions through open and modular architectures, thereby reducing RAN-CN coupling and reconstituting these functions as service-based components~\cite{O-RAN}.

Despite these standardization efforts, current approaches largely remain model-centric, relying on model-specific data pipelines and lifecycle management. As networks scale and diversify, data distribution shifts become more frequent and unpredictable. This requires repeated retraining and redeployment, increasing operational overhead and limiting scalability. More fundamentally, since intelligence is embedded and optimized independently per model, current standardization frameworks provide limited architectural support for coherent RAN-CN coordination.

\textbf{Technical research:} Alongside standardization activities, extensive technical research has investigated how E2E intelligence can be realized in a fully AI-native 6G environment across both architectural and management perspectives.

From an architectural perspective, prior research has proposed AI-native network architectures that move beyond connectivity-centric design to natively support AI model training, inference, and validation within the network~\cite{Zhu25}. The emerging 6G architectures proposed here embed intelligence into the network fabric itself across layers ranging from the physical to the application layer. In addition, Digital Twin (DT)-based approaches have been studied as a means to validate AI-driven control logic and enhance the stability and reliability of intelligent network infrastructures~\cite{Shahid25}.

From a management perspective, existing research has advanced autonomous network control mechanisms that extend zero-touch operations to data-driven cognitive management. Prior studies introduce AI-driven automation that integrates monitoring, analysis, decision-making, and execution. Within these frameworks, service requirements and intents are abstracted into enforceable policies, enabling dynamic orchestration across the RAN and CN. To support this intelligence, several works emphasize integrated E2E data pipelines spanning multiple domains with systematic preprocessing, feature extraction, and normalization. Complementary research further explores collaboration among distributed AI models for scalable network-wide autonomous operation and closed-loop service assurance that monitors service behavior and adapts control policies to correct deviations from intended objectives~\cite{Wang25}.

Despite these advances, existing technical approaches still lack a unified inference framework capable of reasoning about and controlling relationships across both the RAN and CN. Consequently, cross-domain optimization remains inherently fragmented, preventing timely and globally consistent decision-making under dynamic conditions.

\subsection{Background: AI Agent}

\textbf{AI Agent—The Core of Next-Generation Network Intelligence:} As illustrated in Fig. 1, an AI agent goes beyond conventional question-answering systems by acting autonomously to set goals, utilize tools, and solve problems. The AI agent architecture is based on the synergy among its internal components—memory, LLM, the ReAct paradigm, and specialized Tools. The following sections outline the key components that enable AI agents to operate as intelligent decision-makers.

\textbf{Large Language Model—The Agent's Brain:} As the central processing unit shown in Fig. 1, the LLM serves as the core reasoning engine (the "Thought" process in Fig. 1). Although LLMs were originally designed for Natural Language Processing (NLP) tasks, modern LLMs exhibit strong general-purpose reasoning capability. They map heterogeneous observations, objectives, and constraints into coherent internal representations across domains. Rather than relying on fixed input-output mappings learned for a single task, LLMs leverage broad prior knowledge and abstraction capabilities to infer relationships and implications under unseen conditions. These properties position LLMs as the cognitive core for unified inference and decision-making in complex network environments.

\begin{figure}
\centerline{\psfig{figure=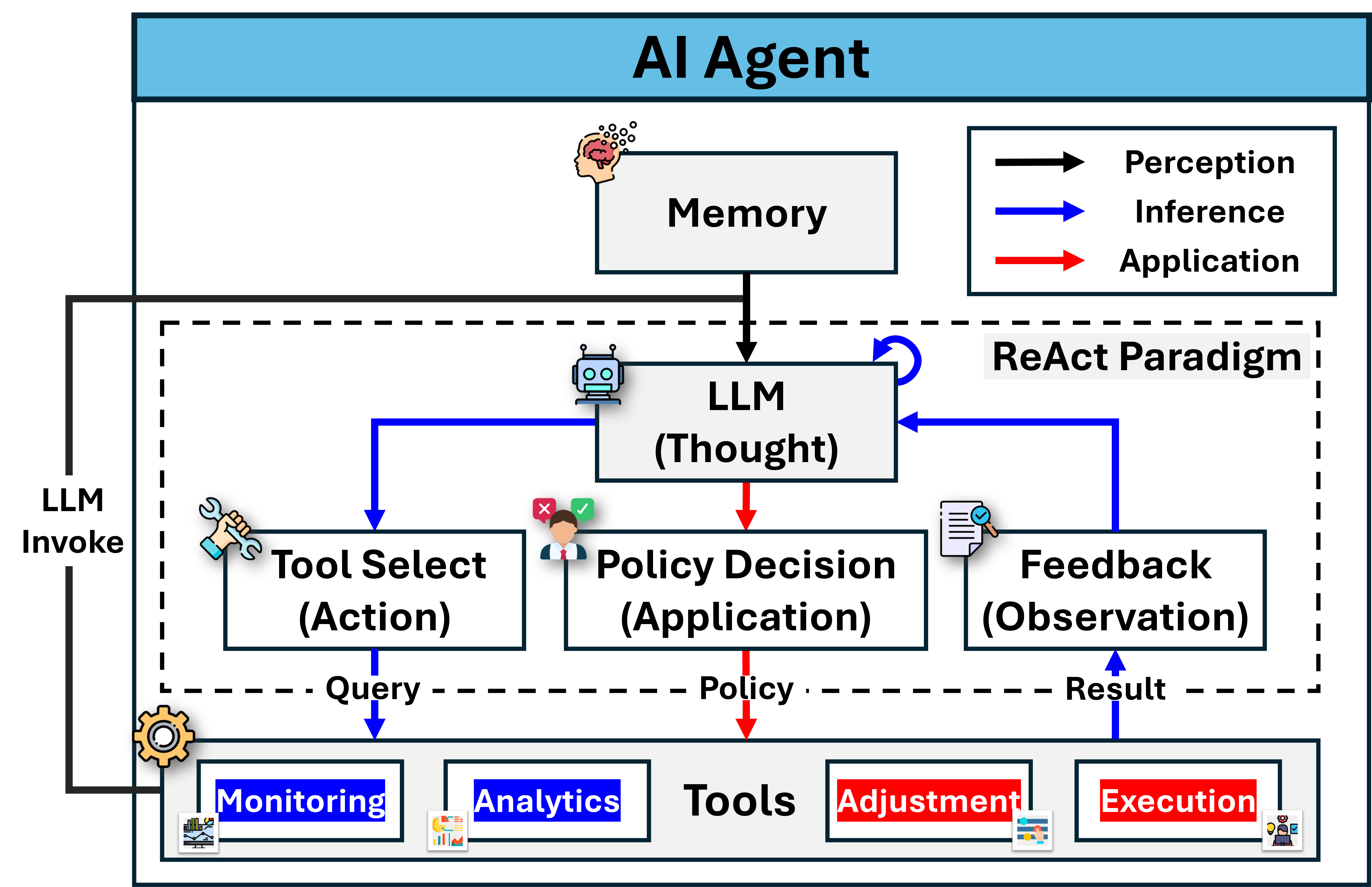,width=9cm}}
\caption{AI Agent with the ReAct paradigm.}
\label{fig:System_model}
\end{figure}

\textbf{ReAct Paradigm—The Agent’s Reasoning–Action Feedback Loop:} The ReAct paradigm transforms LLMs from passive inference engines into interactive reasoning agents capable of engaging with their environment~\cite{ReAct}. As illustrated in the "ReAct Paradigm" block in Fig.~1, ReAct extends one-shot reasoning by interleaving reasoning with actions—such as querying databases, calling APIs, or triggering control functions—and refining decisions based on resulting observations. This thought-action-observation loop establishes a closed feedback structure in which decisions are continuously updated using real-time observations.

Such a feedback-driven reasoning mechanism is well suited to network intelligence, where system states evolve dynamically and cannot be fully captured by static inputs. By enabling the agent to actively acquire additional information and validate intermediate hypotheses through tool-augmented actions, ReAct allows inference to adapt to current network conditions instead of relying solely on pre-trained knowledge. This explicit reasoning-action trace also supports a transparent decision process that can be inspected, logged, and audited, which is important for operational control. These characteristics make ReAct a strong foundation for adaptive and explainable cross-domain inference.

\textbf{Memory—The Agent’s Experience:} To execute complex network control tasks effectively, an AI agent must preserve operational continuity by maintaining awareness of both historical context and evolving system states. As illustrated in Fig.~1, the memory component provides contextual information to the LLM, allowing decisions to incorporate both current observations and prior experiences. 

Memory is structured into short-term and long-term layers to balance immediate responsiveness with sustained learning. Short-term memory leverages the LLM's context window to retain recent reasoning steps, observations, and action outcomes within a session, ensuring coherence. In contrast, long-term memory stores historical failures, prior control actions, and technical artifacts in a vector database, preserving knowledge across sessions. Together, these mechanisms enable the agent to accumulate experience and maintain consistent decision-making under dynamic network conditions.

\textbf{Tools—The Agent’s Hands and Feet:} To perform effective network control, an AI agent must be able to both perceive the operational network state and enact control decisions on external systems. Tools provide this capability by serving as structured interfaces that extend LLM reasoning beyond text-based inference into concrete observation and action. As illustrated in Fig.~1, tools can be categorized into monitoring and analytics functions for perception, and adjustment and execution functions for control. To support scalable and interoperable integration, recent work has emphasized standardized interaction protocols between agents and external systems, such as the Model Context Protocol (MCP)~\cite{MCP}.

While such tool abstractions are broadly applicable to general-purpose AI agents, they are particularly critical for network intelligence. Modern mobile networks comprise heterogeneous and tightly coupled subsystems across the RAN and CN with distinct telemetry, interfaces, and timing constraints. Without explicit tool abstraction, embedding domain-specific logic directly into the LLM would severely limit scalability, adaptability, and maintainability. By externalizing perception and control into well-defined tools, the agent can dynamically acquire real-time network state, invoke control actions across domains, and adapt its decisions as conditions evolve. This separation of reasoning and execution is essential for realizing unified inference and closed-loop control in E2E intelligent network management.

\section{An AI Agent-Based RAN-CN Converged Intelligence Framework}
\label{Sec:System_model}

\begin{figure}
\centerline{\psfig{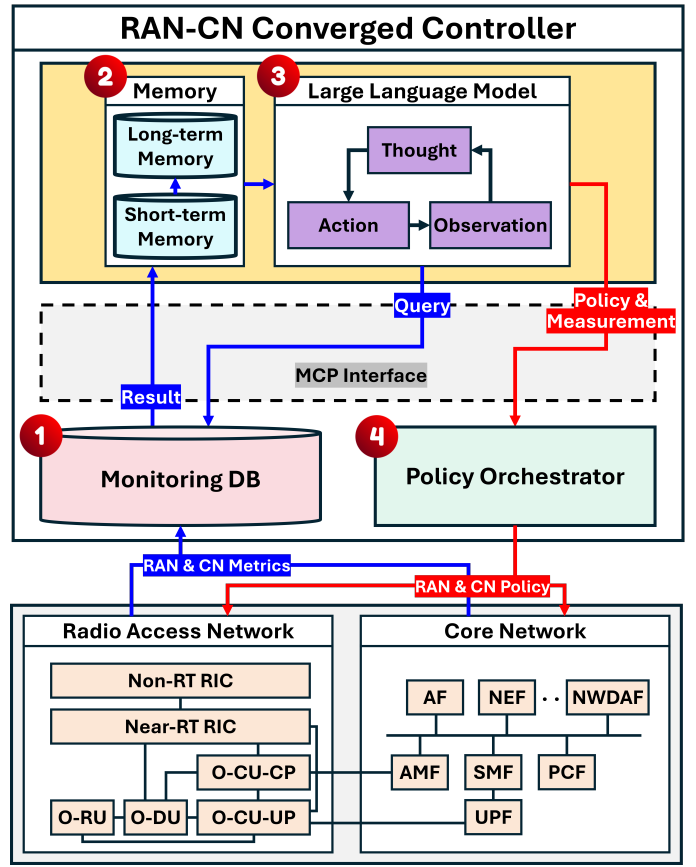}}
\caption{RAN-CN intelligence convergence framework.}
\label{fig:System_model}
\end{figure}

This section presents a novel framework to overcome the limitations of existing AI/ML-based control architectures that operate independently across the RAN and CN. Fig.~2 illustrates the proposed AI agent-based RAN-CN converged intelligence framework, which enables unified inference and coordinated cross-domain control through a central AI agent interacting with distributed network data and control mechanisms.

At the core of the framework is a centralized controller that resides in the management domain and interfaces with both the RAN and CN. It aggregates and interprets cross-domain network state information to derive coordinated control policies. An AI agent based on the ReAct paradigm drives this process through a thought-action-observation loop. The agent first analyzes the current network situation, formulates structured queries to retrieve relevant real-time data from a dedicated monitoring DB, and refines its reasoning based on the observed results. This iterative loop enables dynamic adaptation to network changes. The roles of each component are described below:

The Monitoring DB (labeled (1) in Fig.~2) is a structured data repository that collects real-time network state information from both the RAN and CN. From the RAN side, this includes state and performance indicators such as Physical Resource Block (PRB) utilization, user density, and handover attempt and success rates. From the CN side, the collected information includes UPF processing delay, session counts, slice-level QoS violations, and NWDAF analytics. The collected data are stored in the Monitoring DB in a time-series format and retrieved through MCP-based tool calls. Depending on the data structure and source, SQL or Domain-Specific Languages (DSLs) are used to query and extract data.

The framework incorporates a dual-memory architecture (labeled (2) in Fig.~2) to support context-aware control within the management domain. Short-term memory maintains the working context of each control cycle by caching observations from the monitoring DB, while long-term memory stores accumulated operational knowledge—such as past decisions, failure handling records, and policy references—allowing the agent to ground its decisions in historical network behavior when addressing complex or non-routine scenarios.

Based on the information retrieved from the monitoring DB and memory components, the LLM (labeled (3) in Fig.~2) incrementally constructs hypotheses about the current network state (thought), formulates structured queries to validate or refine its understanding (action), and incorporates the results (observation) to continue its reasoning. This iterative thought-action-observation process enables the LLM to progressively converge toward a consistent interpretation of the network state across the RAN and CN.

Once sufficient confidence in the inferred network state is achieved, the LLM generates high-level control outputs in the form of policies with associated measurement specifications. For example, it may produce directives such as \texttt{ApplyPolicy[PRB\_reservation += 10\%]} or \texttt{ApplyPolicy[Core\_Bandwidth += 1~Gbps]}, along with quantitative indicators—such as PRB utilization, delay distributions, session density, or congestion levels for specific cells or slices—that should be monitored to validate the control decisions. These outputs are not executed directly by the LLM. Instead, they are transmitted via the MCP interface to the policy orchestrator (labeled (4) in Fig.~2), which verifies the commands, translates them into standardized control representations, and dispatches them to enforcement points in the RAN and CN, such as the RIC, NWDAF, or target Network Functions (NFs). This separation of policy generation and execution enhances explainability while ensuring operational safety and controllability.

The proposed framework departs from conventional AI-based control approaches by replacing task-specific input–output models with a unified, reasoning-centric architecture. By enabling LLM-based reasoning over structured real-time data across the RAN and CN, it supports adaptive decision-making without frequent retraining. Moreover, cross-domain awareness allows control decisions to be optimized from an E2E perspective rather than being constrained by isolated domain-specific policies. The explicit thought–action–observation cycle further provides traceable and interpretable decisions suitable for operational control.

\section{Evaluation}
\label{Sec:evaluations}

In this article, we evaluate the proposed framework along three dimensions. First, we examine the LLM's ability to perform unified inference across multiple network tasks using heterogeneous RAN and CN state information. Second, we assess its robustness under previously unseen scenarios, where task-specific ML models often struggle to generalize. Third, we evaluate the E2E optimization capability of the proposed framework. Experiments were conducted on a system equipped with an Intel Core i7-14700K $3.4$~GHz CPU and an NVIDIA RTX~5090 GPU.

\begin{figure*}[!t]
\centerline{\psfig{figure=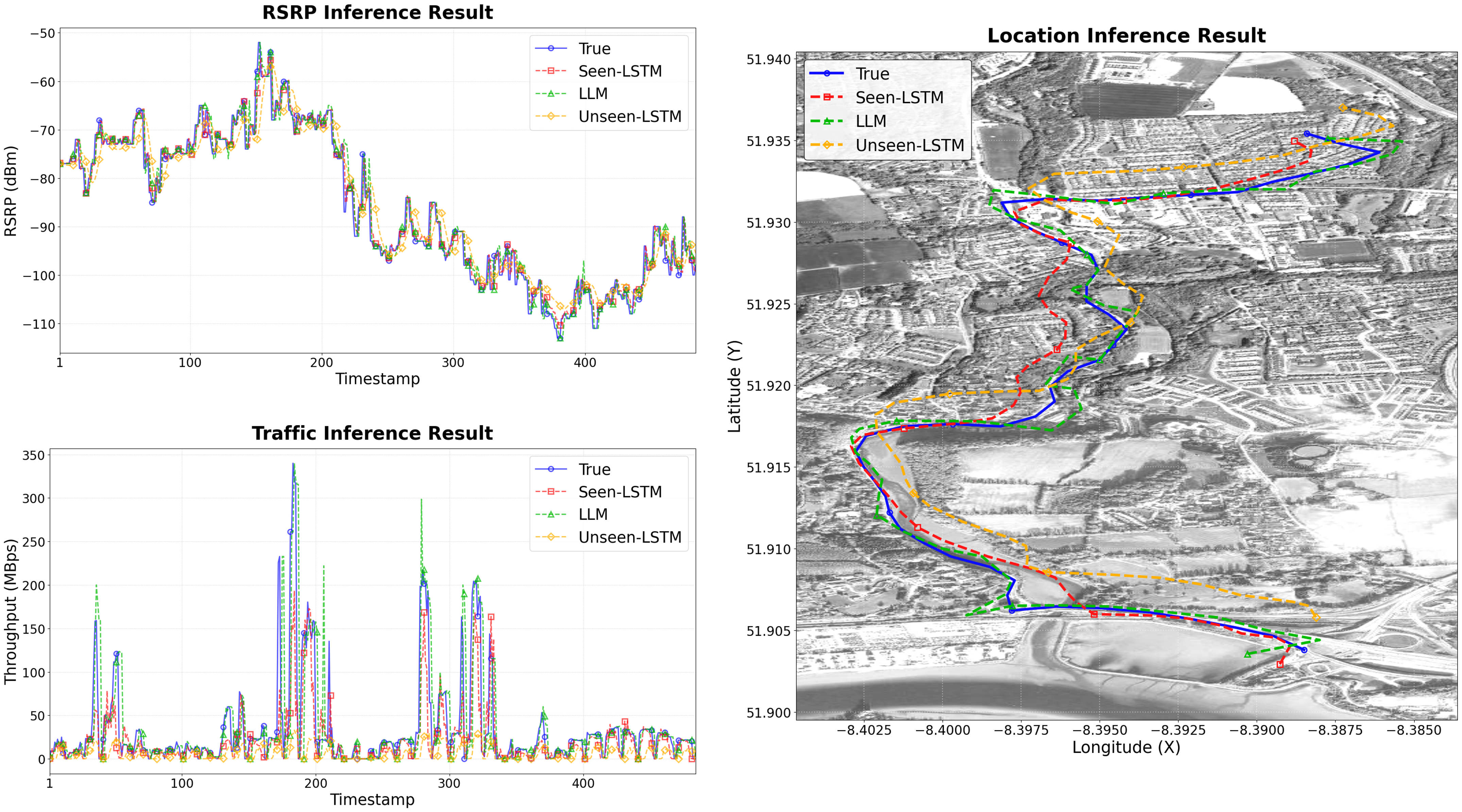,width=18cm}}
\caption{Generalization performance of Seen-LSTM, LLM, and Unseen-LSTM evaluated on RSRP, throughput, and location}
\label{fig:System_model}
\end{figure*}

\begin{table*}[!t]
\centering
\caption{Performance comparison of Seen-LSTM, LLM, and Unseen-LSTM}
\label{tab:performance}

\begin{subtable}[t]{0.32\textwidth}
\centering
\begin{tabular}{lcc}
\toprule
Model & MAE & RMSE \\
\midrule
Seen-LSTM   & 1.7163 & 2.7607 \\
LLM         & 1.7434 & 2.9811 \\
Unseen-LSTM & 3.3938 & 4.4598 \\
\bottomrule
\end{tabular}
\caption{RSRP inference (unit: dBm)}
\end{subtable}
\hfill
\begin{subtable}[t]{0.32\textwidth}
\centering
\begin{tabular}{lcc}
\toprule
Model & MAE & RMSE \\
\midrule
Seen-LSTM   & 16.0950 & 35.5650 \\
LLM         & 18.9251 & 45.0681 \\
Unseen-LSTM & 27.8196 & 54.2901 \\
\bottomrule
\end{tabular}
\caption{Throughput inference (unit: Gbps)}
\end{subtable}
\hfill
\begin{subtable}[t]{0.32\textwidth}
\centering
\begin{tabular}{lcc}
\toprule
Model & MAE & RMSE \\
\midrule
Seen-LSTM   & 0.867 & 0.991 \\
LLM         & 0.492 & 0.645 \\
Unseen-LSTM & 1.115 & 1.199 \\
\bottomrule
\end{tabular}
\caption{Location inference (unit: mdeg)}
\end{subtable}
\end{table*}

\textbf{Unified inference across multiple network tasks:} This evaluation uses the 5G production dataset~\cite{Raca20}, which contains 83 traces from a major Irish mobile operator, covering both stationary and mobile usage patterns across file download and video streaming scenarios. File download traces are used as the test set. The Phi-3-4B-Instruct model serves as the LLM-based agent. For comparison, two Long Short-Term Memory (LSTM) baselines are considered: Seen-LSTM (trained on file download) and Unseen-LSTM (trained on video streaming).

Fig.~3 presents the inference results for heterogeneous network variables, including RSRP, throughput, and user location, while Table~\ref{tab:performance} summarizes the corresponding metrics. Unlike task-specific LSTMs that require separate training for each objective, the LLM performs unified inference across variables. Table~\ref{tab:performance} shows that the LLM achieves performance comparable to Seen-LSTM for RSRP and throughput, and outperforms both baselines in location inference, attaining the lowest MAE and RMSE. These results demonstrate the effectiveness of the proposed approach in addressing the limitations of traditional ML models.

\textbf{Robustness to unseen scenarios:} Using the same experimental setup, we assess the robustness of the LLM-based agent under unseen traffic scenarios against two LSTM baselines. As shown in Fig.~3 and Table~\ref{tab:performance}, Seen-LSTM achieves the highest accuracy when evaluated on scenarios identical to its training data. However, its advantage does not generalize; Unseen-LSTM exhibits substantial performance degradation, with sharp increases in MAE and RMSE across all metrics.

In contrast, the LLM maintains stable performance across both seen and unseen scenarios without additional retraining. This result highlights the inherent limitation of conventional ML models, which rely on the training distribution. By leveraging its general-purpose reasoning capability and broad pre-trained knowledge, the LLM demonstrates stronger robustness to distribution shifts, making it more suitable for dynamic real-world networks.

\textbf{Cross-Domain Optimization:} This evaluation is based on the E2E network slicing problem in~\cite{Testbed}, which maximizes the number of admitted users through joint optimization of RAN radio resources and CN computing capacity. The proposed AI agent-based RAN-CN converged intelligence framework, instantiated with GPT-5-mini, is implemented in a simulated environment. Users are distributed according to a Poisson point process. On the RAN side, large-scale fading is modeled using a log-distance path loss model, while the CN is represented by a fat-tree topology composed of virtual nodes and links. For Service Function Chaining (SFC), eMBB requires 4-6 Virtual Network Functions (VNFs), whereas URLLC requires 2-3 VNFs.

We compared the proposed E2E LLM-based optimization with baselines that differ in how intelligence is distributed across the RAN and CN. The Round Robin (RR) baseline applies static, domain-local resource allocation without learning or adaptation. Domain-specific LLM baselines optimize each domain independently without cross-domain awareness. In contrast, the proposed approach jointly reasons over RAN-CN state information to generate unified allocation decisions, emphasizing the importance of E2E coordination for globally optimized network slicing.

\begin{figure}
\centerline{\psfig{figure=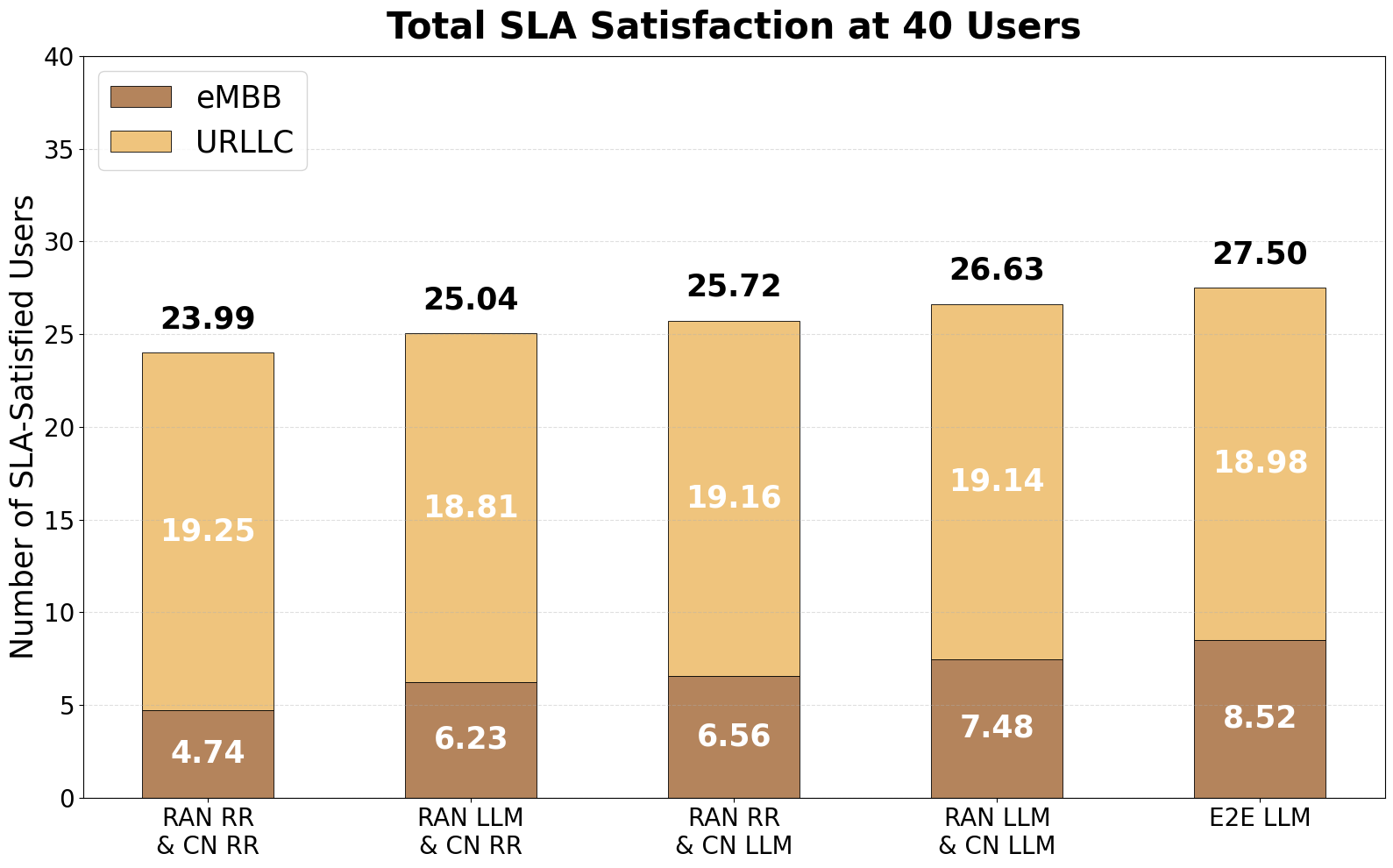,width=9cm}}
\caption{Total SLA satisfaction at 40 users.}
\label{fig:System_model}
\end{figure}

Fig.~4 shows the number of SLA-satisfied users in the 40-user scenario (20 eMBB and 20 URLLC). The proposed E2E LLM-based approach achieves the best performance, satisfying an average of 27.50 users, compared to 26.63 under the combined RAN LLM \& CN LLM strategy. While domain-specific LLMs optimize each domain reasonably, they often converge to domain-wise optimal but E2E suboptimal solutions. For instance, the RAN-side LLM may allocate excessive radio resources to admit more users, but insufficient CN computing capacity prevents satisfying the corresponding E2E SLAs. In contrast, the E2E LLM jointly considers RAN and CN constraints, identifies cross-domain bottlenecks, and prevents inefficient single-domain over-provisioning. By aligning radio and core resources, it consistently achieves higher E2E SLA satisfaction, demonstrating the advantage of unified cross-domain optimization.

\begin{figure}
\centerline{\psfig{figure=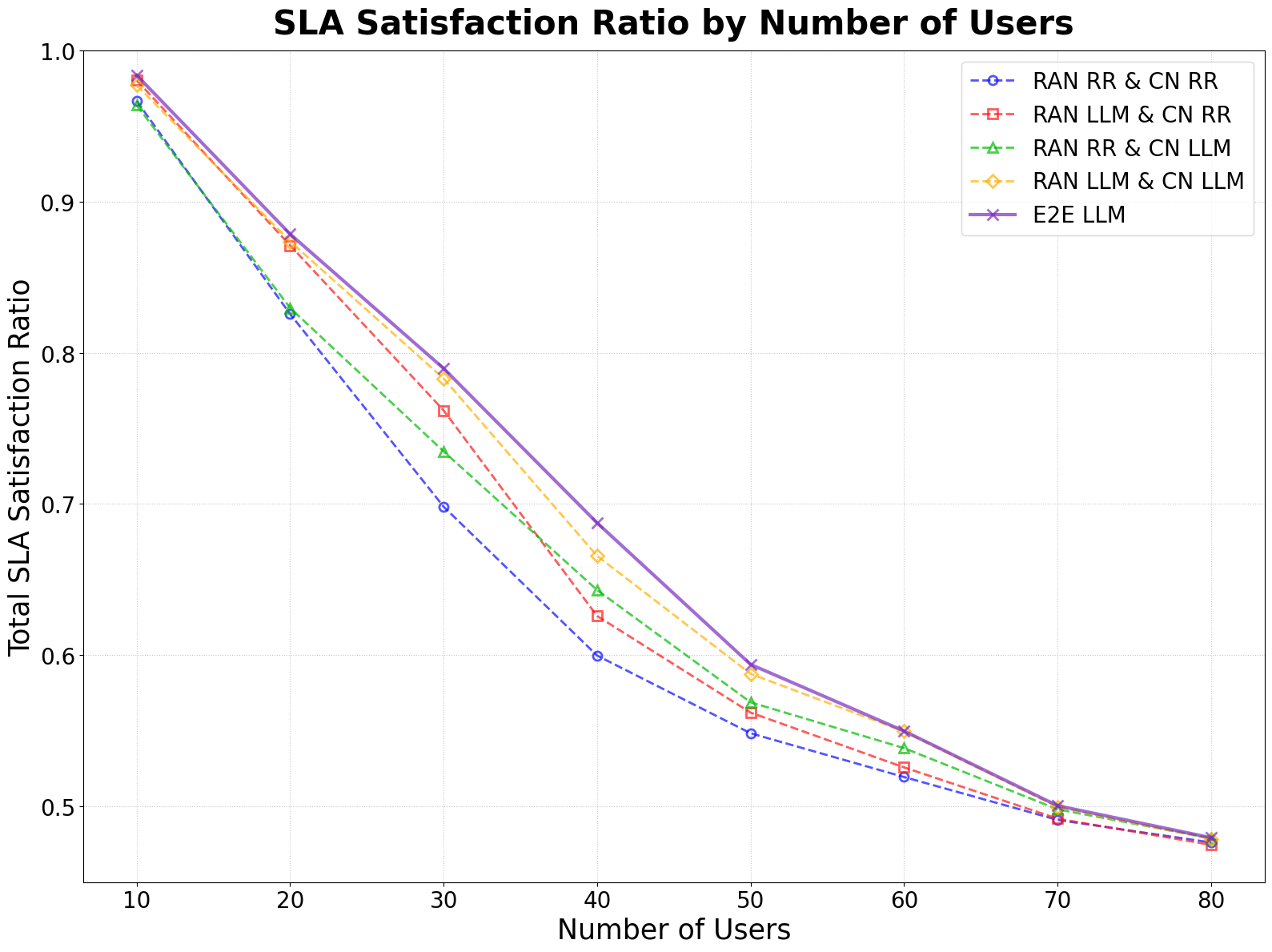,width=9cm}}
\caption{Total SLA satisfaction as a function of the number of users.}
\label{fig:System_model}
\end{figure}

To examine the effectiveness of E2E optimization under varying traffic loads, the number of users is varied from 10 to 80. As shown in Fig.~5, the SLA satisfaction ratio decreases with higher network load due to intensified resource contention in both the RAN and CN. Nevertheless, the proposed E2E LLM-based approach (purple curve) consistently outperforms all baselines across user configurations. This indicates that unified reasoning over coupled RAN-CN constraints enables more effective resource allocation. In contrast, domain-specific approaches that optimize each domain independently show limited coordination, resulting in lower overall efficiency. These results confirm that the proposed framework achieves scalable E2E optimization by aligning radio and core network resources.

\section{Future Research Direction}
\label{Sec:results}

\textbf{Deployment-aware optimization of LLM-based agents:} A key advantage of the proposed framework lies in leveraging the domain-agnostic reasoning capability of LLMs to enable unified inference across the RAN and CN. However, embedding such general-purpose reasoning into real-time control loops introduces latency and computational challenges.

Future research should therefore prioritize optimizing deployment for network control environments. This includes designing control-oriented LLM instantiations that preserve cross-domain reasoning capability while satisfying stringent latency and reliability requirements. Techniques such as instruction tuning with network-specific prompts, model distillation, and structured pruning can realize lightweight real-time implementations, without sacrificing the fundamental generality of the reasoning process.

\textbf{Ensuring safety and verifiability of ReAct-based reasoning:} 
While the ReAct paradigm enables iterative reasoning and action, embedding such agentic behavior into operational network control loops raises safety and reliability concerns. In particular, unconstrained reasoning may trigger inappropriate queries, inconsistent control actions, or policy violations that threaten network stability and service continuity.

Future research must focus on safety- and verification-aware control mechanisms for ReAct-based agents. Promising directions include policy template-based command generation to restrict actions to a predefined safe space, validation of control outputs against domain constraints before execution, and systematic logging of the entire Thought–Action–Observation trajectory. Such mechanisms enable post hoc inspection, auditing, and accountability, forming a foundation for trustworthy LLM-driven network control.

\textbf{Multi-agent cooperative architecture for distributed control:} As network scales, geographical dispersion, and functional heterogeneity increase, a single centralized LLM-based agent may limit scalability, responsiveness, and fault tolerance. Centralized reasoning can introduce latency bottlenecks and single points of failure in large-scale RAN--CN environments.

Future research should investigate multi-agent cooperative architectures, where multiple LLM-based agents are deployed across domains, regions, or functional layers. In such architectures, each agent performs localized reasoning from partial observations while exchanging summarized state information and coordination signals. Through cooperative policy generation and conflict-aware coordination, multi-agent systems can enable scalable, context-aware, and resilient E2E control, extending the proposed framework toward distributed and autonomous 6G.

\textbf{Intelligent data collection and preprocessing pipelines:} The effectiveness of LLM-driven reasoning depends on the quality, freshness, and structural consistency of the input data. In an AI agent-based control framework, inadequate data collection or preprocessing can degrade reasoning accuracy and cause suboptimal or unstable decisions.

Future research should explore intelligent data pipelines integrated with agent-based control loops. Key directions include automated collection of relevant network state information, real-time data ingestion with schema normalization, and adaptive edge-side preprocessing to reduce latency and noise. Such pipelines ensure coherent and reliable state representations in monitoring DB, providing robust and trustworthy E2E reasoning.

\textbf{Prototype development and standard-compliant validation:} To bridge conceptual design and practical deployment, future research should develop a fully functional prototype of the proposed framework that interacts with real network control components such as the Service Management and Orchestration (SMO), RIC, and NWDAF. This prototype is necessary to validate the feasibility of LLM-driven reasoning and ReAct-based control, as well as compatibility with existing operational workflows.

In particular, future efforts should emphasize integration with 3GPP- and O-RAN-compliant interfaces, ensuring seamless operation within standardized control loops. A standards-compliant prototype would allow systematic evaluation of latency, reliability, interoperability, scalability, and robustness in realistic environments. This step is critical for advancing AI agent-based RAN-CN converged intelligence from a research concept toward deployable 6G network solutions.

\section{Conclusion}
\label{Sec:Conclusion}

This work examined the fundamental limitations of existing task-specific AI-based network control approaches that operate independently across the RAN and CN, leading to limited adaptability, high retraining overhead, and suboptimal E2E decisions. To address these challenges, we proposed an AI agent-based RAN-CN converged intelligence framework that integrates LLMs with the ReAct paradigm to enable unified E2E reasoning and closed-loop control without task-specific retraining. Evaluations demonstrate that the proposed framework achieves unified inference across heterogeneous network tasks, robustness under previously unseen conditions, and superior E2E optimization compared to domain-isolated approaches, highlighting the effectiveness of agentic reasoning in handling coupled cross-domain constraints. Overall, this work shifts from fragmented model-centric control to unified reasoning-driven network intelligence, providing a scalable and explainable foundation for autonomous E2E operation in future mobile networks.

\begin{IEEEbiographynophoto}
{YOUBIN HAN} (gksyb4235@khu.ac.kr) received the B.S. degree from the Department of Biomedical Engineering, Kyung Hee National University, Yongin, Korea, in 2025. He is currently M.S student at the Department of Electronics and Information Convergence Engineering, Kyung Hee University, Yongin, Korea. 
\end{IEEEbiographynophoto}

\begin{IEEEbiographynophoto}
{HANEUL KO} [SM'23] (heko@khu.ac.kr) received B.S. and Ph.D. from the School of Electrical Engineering, Korea University, Seoul, Korea, in 2011 and 2016, respectively. He is currently an associate professor in the Department of Electronic Engineering, Kyung Hee University, Yongin, Korea.
\end{IEEEbiographynophoto}

\begin{IEEEbiographynophoto}
{NAMSEOK KO} (nsko@etri.re.kr) received the M.S. and Ph.D. degrees from KAIST, Korea, in 2000 and 2015, respectively.  In 2000, he joined Electronics and Telecommunications Research Institute (ETRI), Korea. He is currently a Director at the Mobile Core Network Research Section in ETRI. As of 2023, he is also an associate professor at the Department of Information and Communication Engineering at the University of Science and Technology (UST).
\end{IEEEbiographynophoto}

\begin{IEEEbiographynophoto}
{Tarik Taleb} [SM'23] (Tarik.Taleb@ruhr-uni-bochum.de) received the B.E. degree (with distinction) in information engineering, and the M.Sc. and Ph.D. degrees in in formation sciences from Tohoku University, Sendai, Japan, in 2001, 2003, and 2005, respectively. He is currently a full professor with Ruhr University Bochum, Bochum, Germany. 
\end{IEEEbiographynophoto}

\begin{IEEEbiographynophoto}
{Yan Chen} (yanchen@ieee.org) received the B.E. degree in Information Engineering and the Ph.D. degree in Information and Communication Engineering from China University of Mining and Technology, China, in
2016 and 2022. He is currently a Postdoctoral Researcher with Ruhr University Bochum, Bochum, Germany.
\end{IEEEbiographynophoto}

\end{document}